\documentclass[final]{svjour2}
\usepackage{graphicx}
\usepackage{rotating}
\usepackage{amssymb}
\usepackage{mathptmx}
\usepackage[numbers]{natbib}
\makeatletter
\journalname{Journal of Low Temperature Physics}

\bibpunct{}{}{,}{s}{}{,}

\begin{document}

\def\ca{$\sim$}

\newcommand{\hdblarrow}{H\makebox[0.9ex][l]{$\downdownarrows$}-}
\title{Pulse Shape Analysis with scintillating bolometers}

\author{L. Gironi $^1$}

\institute{1:Dipartimento di Fisica, Universit\`a degli Studi di Milano - Bicocca,\\ Milano, 20126, IT\\
Tel.:+39 02 64 48 21 07 \\ Fax:+39 02 64 48 24 63\\
\email{luca.gironi@mib.infn.it}
}

\date{01.08.2011}

\maketitle

\keywords{Bolometers, Scintillators, Pulse shape discrimination }

\begin{abstract}

Among the detectors used for rare event searches, such as neutrinoless Double Beta Decay (0$\nu$DBD) and Dark Matter experiments, bolometers are very promising because of their favorable properties (excellent energy resolution, high detector efficiency, a wide choice of different materials used as absorber, ...). However, up to now, the actual interesting possibility to identify the interacting particle, and thus to greatly reduce the background, can be fulfilled only with a double read-out (i.e. the simultaneous and independent read out of heat and scintillation light or heat and ionization). This double read-out could greatly complicate the assembly of a huge, multi-detector array, such as CUORE and EURECA. The possibility to recognize the interacting particle through the shape of the thermal pulse is then clearly a very interesting opportunity.

While detailed analyses of the signal time development in purely thermal detectors have not produced so far interesting results, similar analyses on macro-bolometers ($\sim$10-500 g) built with scintillating crystals showed that it is possible to distinguish between an electron or $\gamma$-ray and an $\alpha$ particle interaction (i.e. the main source of background for 0$\nu$DBD experiments based on the bolometric technique). Results on pulse shape analysis of a CaMoO$_4$ crystal operated as bolometer is reported as an example. An explanation of this behavior, based on the energy partition in the heat and scintillation channels, is also presented.

PACS numbers: 23.40B, 95.35.+d, 07.57.K, 29.40M, 66.70.-f 
\end{abstract}

\section{Introduction}

In neutrinoless Double Beta Decay (0$\nu$DBD) \cite{BBreview}, as in all the rare event studies, spurious events are a limiting factor to the reachable sensitivity of the experiment. Unfortunately natural radioactive background is often present in the detector itself or in the materials surrounding it, no matter how much one can try to reduce it with shieldings. In order to reduce this residual unavoidable background, all the possible approaches require both a good energy resolution and the capability to identify the nature of the projectile that interacted with the detector.

Bolometers \cite{bolometers} are based on the detection of phonons produced after an energy release by an interacting particle and can have both an excellent energy resolutions and extremely low energy thresholds with respect to conventional detectors. They can be fabricated from a wide variety of materials, provided they have a low enough heat capacity at low temperatures. The latter is a priceless feature for experiments that aim at detectors containing particular atomic or nuclear species to optimize the detection efficiency. If other excitations (such as ionization charge carriers or scintillation photons) are collected in addition to phonons, bolometers have already shown to be able to discriminate nuclear recoils from electron recoils, or $\alpha$ particles from $\beta$ particles and $\gamma$-rays. Recently has been demonstrated \cite{psa} the possibility to obtain similar results by Pulse Shape Analysis (PSA), without the requirement of a double readout for phonons and ionization or scintillation light.  

\section{Scintillating Bolometers and Pulse Shape Analysis}
\label{BolTec}

The concept of a scintillating bolometer is very simple: a bolometer coupled to a light detector \cite{CaF2}. The first must consist of a scintillating absorber thermally linked to a phonon sensor while the latter can be any device able to measure the emitted photons. Excellent results were obtained using as light detector a second `dark' thin bolometer \cite{LD}. In this device the simultaneous and independent read out of the heat and the scintillation light permits to discriminate events due to $\beta$/$\gamma$, $\alpha$ and neutrons thanks to their different scintillation yield.
Dark Matter as well as 0$\nu$DBD searches can benefit of this capability of tagging the different particles, and more generally this technique can be exploited in any research where background suppression or identification is important.

The development of such a hybrid detector, able to discriminate $\alpha$ particles, was the main purpose of studies on scintillating bolometers for 0$\nu$DBD searches. Several devices\cite{CristalliTest}, differing mainly in the scintillating crystal material and size, were tested to study their thermal response, light yield and radio purity. In these studies was discovered an extremely interesting feature of some of the tested crystals: the different pulse shape of the thermal signals produced by $\alpha$'s and by $\beta$/$\gamma$'s. This feature opens the possibility of realizing a bolometric experiment that can discriminate among different particles, without the need of a light detector coupled to each bolometer. In the case of a huge, multi-detector array, such as CUORE \cite{CUORE} and EURECA \cite{EURECA}, the benefits of employing this technique would be impressive:

\begin{itemize}
\item 
more ease during assembly because the single element of the array would be a quite simpler device and light collectors are no more needed.
\item
fewer readout channels, with an evident reduction of cost and work and also a cryogenic benefit: a significant reduction of the heat load of the readout channels.
\item
a significant cost reduction, saving money and work that would be necessary for the light detectors procurement and optimization. 
\end{itemize}

Up to now, the discrimination based on PSA has been very effective for particles with energy higher than few hundreds of keV. This technique can therefore already be applied to 0$\nu$DBD experiments while can not yet be applied to study low-energy events as required for Dark Matter experiments. However, since these are the first observations of this feature, there are considerable rooms for improvement, for example by using phonon sensors and electronics better suited to this type of measurements than those used (Ge-NTD thermistors). This could significantly lower the energy threshold of the pulse shape discrimination.

Finally, the removal of the light guide - used in scintillating bolometer to collect more light - allows to face these devices to the assembly materials by opening the possibility to study surface contaminations of these materials with higher sensitivity than standard technique (e.g. Si surface barrier detectors). The study of such contamination is very interesting because they are indicated as the main source of background for 0$\nu$DBD experiments based on the bolometric technique\cite{FondoCuoricino}.

\section{Experimental Results}

Several experimental evidences\cite{psa} have shown the possibility to discriminate the type of interacting particles through different shape of the thermal signal measured in different scintillating materials (CaMoO$_{4}$, MgMoO$_{4}$, ZnMoO$_{4}$, ZnSe). As an example, I will report here results obtained with a CaMoO$_{4}$ crystal with a mass of 158~g (h = 40mm, $\varnothing$ = 35mm) operated as bolometer in a dilution ($^3$He/$^4$He) refrigerator installed in the hall C of the Underground Laboratory of Gran Sasso.

\begin{figure}[th]
\begin{center}
\includegraphics[ width=0.7\linewidth]{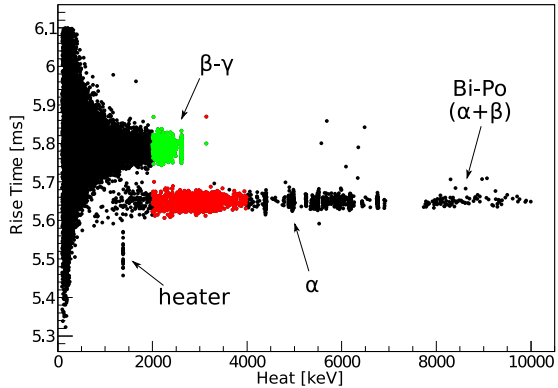}
\end{center}
\caption{(Color online) Scatter plot of $\tau_{rise}$ vs. Heat in CaMoO$_{4}$ crystal. In green (light in the upper curve) events in the 2-4 MeV region due to $\beta$/$\gamma$ particles and in red $\alpha$ events. The energy calibration is performed on the $\gamma$ peaks. The total live time of this measurement was $\sim$43 h. In order to have a high number of $\alpha$ counts in the 2-4 MeV region in these measurements a degraded $^{238}$U source\cite{CristalliTest} was placed in front of the crystal.}
\label{fig:ScatterCaMoO4}
\end{figure}

In figure \ref{fig:ScatterCaMoO4} is reported the thermal rise time $\tau_{rise}$ (evaluated on the recorded raw-pulse as (t$_{90\%}$-t$_{10\%}$)) as a function of the energy released in the bolometer. Two separate bands, ascribed to $\beta$/$\gamma$ and $\alpha$ events can be identified. With different color are reported $\beta$/$\gamma$ and $\alpha$ events having an energy between 2 and 4 MeV where are expected to attend the peaks of the most promising 0$\nu$DBD isotopes. A gaussian fit of these two distribution yields a rise-time of (5.788$\pm$0.017)~ms for $\beta$/$\gamma$'s and of (5.649$\pm$0.013)~ms for $\alpha$'s resulting in a discrimination confidence level\footnote{The discrimination confidence level D$_{Rise}$ is defined as $ D_{Rise} = \frac{Rise_{\beta/\gamma}-Rise_{\alpha}}{\sqrt{\sigma_{\beta/\gamma}^2+\sigma_{\alpha}^2}} $ where $Rise_{\beta/\gamma}$ and $Rise_{\alpha}$ are the average rise times produced by the two kinds of particles  and $\sigma_{\beta/\gamma}$ and $\sigma_{\alpha}$ is the width of the two distributions. } of D$_{Rise}$=6.5$\sigma$.

Similar results were obtained with other crystals (MgMoO$_{4}$, ZnMoO$_{4}$, ZnSe) showing that this is a common feature of some scintillating crystals. Instead, up to now, no evidence of this feature was observed in purely thermal detector.

The different shape of thermal signal produced by $\alpha$ particles and $\beta$/$\gamma$ particles could be explained by the dependence of light yield on the nature of the interacting particle. The high ionization density of $\alpha$ particles implies that all the scintillation states along their path are occupied\cite{BIRKS}. This saturation effect does not occur or at least is much less for $\beta$/$\gamma$ particles. Therefore, in $\alpha$ interactions a larger fraction of energy flow in the heat channel with respect to $\beta$/$\gamma$ events. This leads not only to a different light and heat yield but also to a different time evolution of both signals. The pulse shape of the thermal signal then can be explained by the partition of energy in the two channels. 

\section{The energy partition and the PSA}
\label{sec:ThermalModel}

A possible interpretation of the pulse shape discrimination observed in some scintillating bolometers is based on the following assumption: the processes that lead to the thermal pulse formation (i.e. the production of scintillation light and phonons and their collection in the sensor) can not be considered instantaneous even in slow detectors such as the bolometers. Indeed, at low temperature the averaged scintillation decay time of molybdates\cite{TimeDep1,TimeDep2} is of the order of hundreds of $\mu$s and therefore it is comparable to the typical rise time of the heat signal of our scintillating bolometers.

In the following a possible simplified model to explain the pulse shape discrimination observed in scintillating bolometers will be presented. Let us assume that the interaction of a $\beta$ particle with the crystal produces N$_\beta$ electron/hole pairs. A fraction of these ($\omega_\beta$) will recombine through scintillation processes while the remaining (1-$\omega_\beta$) will give rise to thermal processes (this holds in the case in which only two channels share the total energy). For an $\alpha$ particle the fraction that will recombine through scintillation processes ($\omega_\alpha$) will be lower ($\omega_\alpha < \omega_\beta$) because of saturation effects. In fact, most scintillating centers are occupied and a larger fraction of energy flow through the heat channel. 

Since the thermal processes have a very fast time evolution ($<10^{-8}$s) they can be considered instantaneous with respect to thermal pulse in macro-bolometers. Therefore, let us analyze more in detail the scintillation channel.

\begin{figure}[th]
\begin{center}
\includegraphics[ width=0.5\linewidth]{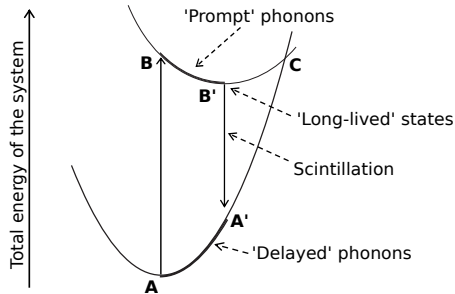}
\end{center}
\caption{Schematic configuration coordinate diagram of the scintillation process. The energy E is plotted versus the coordinate Q (configurational coordinate in the lattice).}
\label{fig:Scint}
\end{figure}

In figure \ref{fig:Scint} is shown a schematic representation of the scintillation channel. It could be divided in 3 stages. In the first stage (B-B$'$), which includes from the initial energy deposition to the formation of excited luminescent centers, there is the emission of `prompt' phonons with typical time $<10^{-8}$s \cite{Lecoq}. These phonons are then emitted with those of the heat channel and do not give any information regarding the type of particle that has released energy. The first stage is followed by the emission of scintillation light (B$'$-A$'$).
This process is very slow if compared to the production of phonons. For instance the measured scintillation decay time constant for CaMoO$_4$ is $380\pm20$ $\mu$s at 7K\cite{CaMoO4}. The scintillation process is then followed by a further stage of non-radiative de-excitation (A$'$-A). Phonons emitted from this stage are then produced at a later time (i.e. they are produced with an exponential behavior equal to the decay time of scintillation) with respect to the `prompt' phonons. 

The fraction $\eta$ of the scintillation channel that gives rise to non-radiative processes is not negligible. For some molybdates \cite{CaMoO4,MgMoO4} it could be as high as $\sim$50\% of the energy that flow in the scintillation channel. Taking into account that the light yield of molybdates at low temperature is higher than tens of keV/MeV \cite{CaMoO4}, this fraction could modify the total thermal pulse shape.
In summary the development of the total thermal signal caused by the interaction of a $\beta$ particle will be due the sum of the heat channel  ($\omega_\beta$) and the non-radiative part of the scintillation channel $\eta$(1 - $\omega_\beta $) that is partly emitted later. 

In the case of the interaction of an $\alpha$ particle there will be an higher percentage of non-radiative processes ($\omega_\alpha > \omega_\beta$) resulting in a different time evolution of the total thermal process. The rise and the decay time for $\alpha$ events measured with the CaMoO4 crystal are $\sim$200 $\mu$s faster than those of $\beta$/$\gamma$ events.

Based on this model, the observation of this effect is therefore related to the combination of two factors: the scintillation decay time (which should be long enough) and the percentage of non-radiative de-excitation of the scintillation channel. In addition, the processes of self-absorption should add up to the effect just described. Therefore, it is expected a dependence on the shape and size of the crystal.

Finally, the processes described above relate only to the energy subdivision and to the time evolution of the heat and light channels. The heat channel induces the temperature rise of the scintillating crystal while the restoration of the equilibrium temperature is mainly due to the thermal link to the heat bath. Further studies are necessary in order to include this simplified model in a more general thermal model of scintillating bolometer that could be able to describe the measured thermal pulse shape.

\section{Conclusion}

The possibility to discriminate the nature of the particle interacting in a bolometric detector, simply on the base of the shape of the thermal pulse, is proved and opens new possibilities for the application of these devices in the field of rare events searches. A possible explanation of this behavior based on the energy partition in two channels (heat and light) is presented. Further studies are therefore being developed to analyze the convolution of the phonons production process and the detector thermal response.

\end{document}